\documentclass[aps,prl,showpacs,twocolumn,lineno,superscriptaddress]{revtex4}  
\usepackage{graphicx}  
\usepackage{dcolumn}   
\usepackage{bm}        
\usepackage{amssymb}   

\newcommand{\MET}{\rlap{\kern0.25em/}E_T}
\begin{document}

\title{Search for Second-Generation Scalar Leptoquarks in
$\bm{p \bar{p}}$ Collisions at $\sqrt{s}$=1.96 TeV\\ }
\affiliation{Institute of Physics, Academia Sinica, Taipei, Taiwan 11529, Republic of China} 
\affiliation{Argonne National Laboratory, Argonne, Illinois 60439} 
\affiliation{Institut de Fisica d'Altes Energies, Universitat Autonoma de Barcelona, E-08193, Bellaterra (Barcelona), Spain} 
\affiliation{Baylor University, Waco, Texas  76798} 
\affiliation{Istituto Nazionale di Fisica Nucleare, University of Bologna, I-40127 Bologna, Italy} 
\affiliation{Brandeis University, Waltham, Massachusetts 02254} 
\affiliation{University of California, Davis, Davis, California  95616} 
\affiliation{University of California, Los Angeles, Los Angeles, California  90024} 
\affiliation{University of California, San Diego, La Jolla, California  92093} 
\affiliation{University of California, Santa Barbara, Santa Barbara, California 93106} 
\affiliation{Instituto de Fisica de Cantabria, CSIC-University of Cantabria, 39005 Santander, Spain} 
\affiliation{Carnegie Mellon University, Pittsburgh, PA  15213} 
\affiliation{Enrico Fermi Institute, University of Chicago, Chicago, Illinois 60637} 
\affiliation{Joint Institute for Nuclear Research, RU-141980 Dubna, Russia} 
\affiliation{Duke University, Durham, North Carolina  27708} 
\affiliation{Fermi National Accelerator Laboratory, Batavia, Illinois 60510} 
\affiliation{University of Florida, Gainesville, Florida  32611} 
\affiliation{Laboratori Nazionali di Frascati, Istituto Nazionale di Fisica Nucleare, I-00044 Frascati, Italy} 
\affiliation{University of Geneva, CH-1211 Geneva 4, Switzerland} 
\affiliation{Glasgow University, Glasgow G12 8QQ, United Kingdom} 
\affiliation{Harvard University, Cambridge, Massachusetts 02138} 
\affiliation{Division of High Energy Physics, Department of Physics, University of Helsinki and Helsinki Institute of Physics, FIN-00014, Helsinki, Finland} 
\affiliation{University of Illinois, Urbana, Illinois 61801} 
\affiliation{The Johns Hopkins University, Baltimore, Maryland 21218} 
\affiliation{Institut f\"{u}r Experimentelle Kernphysik, Universit\"{a}t Karlsruhe, 76128 Karlsruhe, Germany} 
\affiliation{High Energy Accelerator Research Organization (KEK), Tsukuba, Ibaraki 305, Japan} 
\affiliation{Center for High Energy Physics: Kyungpook National University, Taegu 702-701; Seoul National University, Seoul 151-742; and SungKyunKwan University, Suwon 440-746; Korea} 
\affiliation{Ernest Orlando Lawrence Berkeley National Laboratory, Berkeley, California 94720} 
\affiliation{University of Liverpool, Liverpool L69 7ZE, United Kingdom} 
\affiliation{University College London, London WC1E 6BT, United Kingdom} 
\affiliation{Massachusetts Institute of Technology, Cambridge, Massachusetts  02139} 
\affiliation{Institute of Particle Physics: McGill University, Montr\'{e}al, Canada H3A~2T8; and University of Toronto, Toronto, Canada M5S~1A7} 
\affiliation{University of Michigan, Ann Arbor, Michigan 48109} 
\affiliation{Michigan State University, East Lansing, Michigan  48824} 
\affiliation{Institution for Theoretical and Experimental Physics, ITEP, Moscow 117259, Russia} 
\affiliation{University of New Mexico, Albuquerque, New Mexico 87131} 
\affiliation{Northwestern University, Evanston, Illinois  60208} 
\affiliation{The Ohio State University, Columbus, Ohio  43210} 
\affiliation{Okayama University, Okayama 700-8530, Japan} 
\affiliation{Osaka City University, Osaka 588, Japan} 
\affiliation{University of Oxford, Oxford OX1 3RH, United Kingdom} 
\affiliation{University of Padova, Istituto Nazionale di Fisica Nucleare, Sezione di Padova-Trento, I-35131 Padova, Italy} 
\affiliation{LPNHE-Universite de Paris 6/IN2P3-CNRS} 
\affiliation{University of Pennsylvania, Philadelphia, Pennsylvania 19104} 
\affiliation{Istituto Nazionale di Fisica Nucleare Pisa, Universities of Pisa, Siena and Scuola Normale Superiore, I-56127 Pisa, Italy} 
\affiliation{University of Pittsburgh, Pittsburgh, Pennsylvania 15260} 
\affiliation{Purdue University, West Lafayette, Indiana 47907} 
\affiliation{University of Rochester, Rochester, New York 14627} 
\affiliation{The Rockefeller University, New York, New York 10021} 
\affiliation{Istituto Nazionale di Fisica Nucleare, Sezione di Roma 1, University of Rome ``La Sapienza," I-00185 Roma, Italy} 
\affiliation{Rutgers University, Piscataway, New Jersey 08855} 
\affiliation{Texas A\&M University, College Station, Texas 77843} 
\affiliation{Istituto Nazionale di Fisica Nucleare, University of Trieste/\ Udine, Italy} 
\affiliation{University of Tsukuba, Tsukuba, Ibaraki 305, Japan} 
\affiliation{Tufts University, Medford, Massachusetts 02155} 
\affiliation{Waseda University, Tokyo 169, Japan} 
\affiliation{Wayne State University, Detroit, Michigan  48201} 
\affiliation{University of Wisconsin, Madison, Wisconsin 53706} 
\affiliation{Yale University, New Haven, Connecticut 06520} 
\author{A.~Abulencia}
\affiliation{University of Illinois, Urbana, Illinois 61801}
\author{D.~Acosta}
\affiliation{University of Florida, Gainesville, Florida  32611}
\author{J.~Adelman}
\affiliation{Enrico Fermi Institute, University of Chicago, Chicago, Illinois 60637}
\author{T.~Affolder}
\affiliation{University of California, Santa Barbara, Santa Barbara, California 93106}
\author{T.~Akimoto}
\affiliation{University of Tsukuba, Tsukuba, Ibaraki 305, Japan}
\author{M.G.~Albrow}
\affiliation{Fermi National Accelerator Laboratory, Batavia, Illinois 60510}
\author{D.~Ambrose}
\affiliation{Fermi National Accelerator Laboratory, Batavia, Illinois 60510}
\author{S.~Amerio}
\affiliation{University of Padova, Istituto Nazionale di Fisica Nucleare, Sezione di Padova-Trento, I-35131 Padova, Italy}
\author{D.~Amidei}
\affiliation{University of Michigan, Ann Arbor, Michigan 48109}
\author{A.~Anastassov}
\affiliation{Rutgers University, Piscataway, New Jersey 08855}
% ...
\collaboration{CDF Collaboration}
\noaffiliation
\date{\today}

\begin{abstract}

Results on a search for pair production of second generation scalar leptoquark in 
 $p \bar{p}$ collisions at $\sqrt{s}$=1.96 TeV are reported.
The data analyzed were collected by the CDF detector during the 2002-2003 Tevatron Run II
and correspond to an integrated luminosity of 198 pb$^{-1}$.
Leptoquarks (LQ) are sought through their decay into (charged) leptons and quarks, with final state
signatures represented by two muons and jets and one muon, large transverse missing energy and jets.
We observe no evidence for $LQ$ production
and derive 95$\%$ C.L. upper limits on the $LQ$ production cross sections as well as lower limits on 
their mass  as a function of $\beta$, where $\beta$ is the branching fraction for $LQ \rightarrow \mu q$.

\end{abstract}


\maketitle

\newpage


The symmetry between leptons and quarks present in the 
standard model (SM) of particle physics, has lead to the theoretical speculation that a more
foundamental force could operate at energy scales larger than the electroweak 
symmetry breaking scale, allowing for quark to lepton transitions, mediated by new gauge bosons.
Theories like grand unification or R-parity violating supersymmetric models
introduce the idea of quark to lepton transitions\cite{models}. Whenever quarks and leptons
are allowed to couple directly to each other, new bosons carrying both lepton
and baryon quantum numbers can also exist. They are called leptoquark (LQ)\cite{LQ},
they can have spin 0 (scalar LQ) or 1 (vector  LQ),are color-triplet particles, and can either be produced singly 
or in pairs. Most of the other characteristics, such as weak ispospin, electric charge and 
the coupling $\lambda$ to lepton and quark are model dependent\cite{miriam}. Their masses are not predicted.
To accomodate experimental constraints on flavor changing neutral currents, LQ
are assumed to couple to fermions of the same generation\cite{miriam,newtheory}. 
While first generation LQ have been extensively searched for at $e-p$, $e^+e^-$ and $p\bar p$ 
colliders, via single and pair production, and strong limits on their production cross section and mass
have been set\cite{firstGen,newTeV}, second and third generation LQ are 
detectable via pair production at 
hadronic colliders and can be singly produced at $e^+e^-$ machines. Indeed upper limits on second generation 
LQ production cross section and lower limits on their masses exist from LEP\cite{d03}
and the Tevatron Run I\cite{tevatron,SecGenTeV1}, 
but they are usually weaker than the ones for first generation LQ.
In this paper we present new results on a search for 
second generation LQ pair production in $p \bar{p}$ collisions at $\sqrt{s}$=1.96 TeV. 
Assuming two different values for the LQ branching fraction $\beta$ to muon and quark,
we consider the following final state signatures: both LQ decaying 
into muons of opposite charge and quarks ($\beta = 1$), and 
one of the LQ decaying into a muon and a quark and  the other into a 
neutrino and a quark ($\beta = 0.5$). 
Since we do not observe evidence for LQ production, we set
an upper limit on the production cross section times branching fraction (Br) where
Br = $\beta^{2}$ or Br = $2\beta(1-\beta)$.
These results are then combined with the one from a search for  scalar LQ pairs decaying into 
$\nu\nu qq$, resulting in jets and missing transverse energy topology\cite{prl7200}. In this way we can 
express our limits as a continous function of the parameter $\beta$.

CDF is a general--purpose detector built to study
the physics of $p\bar{p}$ collisions at the Tevatron accelerator at Fermilab and it is
described in detail elsewhere \cite{CDF4}.
We use a cylindrical coordinate system around the beampipe in which 
$\theta$ is the polar angle, $\phi$ is the azimuthal angle and $\eta = -{\rm ln} (\rm{tan} ({\theta\over 2})) $.
The transverse energy is defined as  
$E_T  = E \sin \theta $ and the transverse momentum is defined as
$P_T =  P \sin \theta $, where $E$ is the energy measured by the calorimeter and 
$P$ the momentum measured by the tracking system. 
The vector $\bar { \MET}$  is equal to $ -\Sigma E^i_T \bar {n_i}$ 
where $\bar {n_i}$ 
is a unit vector that points from the interaction vertex to the {\it i}th calorimeter tower in the 
transverse plane. 
Its magnitude, $\MET$, is called transverse missing energy. $\MET$ is corrected following the correction of jet energies, and
if muons are identified in the event, $\MET$ is corrected for the muon momenta. 

The data used in the analysis were collected during the 2002-2003 Tevatron Run II. The integrated luminosity
for this data sample is 198  $\pm$ 12 pb$^{-1}$. Events are selected online by requiring
track segments (``stubs'') reconstructed in the muon 
chambers and matched to individual tracks reconstructed in the central tracker (high $P_T$ muon trigger).
The efficiency of the trigger combinations used in the $\mu\mu jj$ and $\mu \nu jj$ analyses,
measured using $Z\to \mu^+ \mu^-$ data\cite{zpaper,muge}, is $\sim$90\%, varying from about 87\% to 95\%
depending on the type of muon chamber used to detect the candidate muon.
Muons are selected as ``tight'' or ``loose'' 
(this second category being used in the $\mu \mu$ analysis only).
A tight muon requires a reconstructed track segment in the muon chambers with positions well matched to the 
extrapolation of a single track, while a loose muon is the one selected by 
requiring only one isolated track. In both cases the energy deposition in the calorimeters
must be consistent with that of  a minimum-ionizing particle.
We apply a cut on the $\chi^2$ of the track fit to eliminate kaons and pions which 
have decayed in flight. 
The identification efficiency for muons has also been measured using data\cite{zpaper,muge} and is 
approximately $90\%$, going from 89\% to 95\% for different types of detector and selection used
to identify the candidate muon. 
The coordinate of the lepton (also assumed to be the event coordinate)
along the beamline must fall within $60$ cm of the center of the 
detector ( $z_{vertex}$ cut) to ensure a good energy measurement in the calorimeter. 
This cut has an efficiency of $(95 \pm 0.1 (stat) \pm 0.5(sys))$\%, 
and is determined from studies of minimum bias events. 
The efficiencies of the identification cuts, the trigger selection and the vertex cut,  
measured using data are taken into account by using scale factors between data and
Monte Carlo events.
Jets are reconstructed using a cone of fixed radius 
$R = \sqrt{(\Delta\eta^2 + \Delta \phi^2)} = 0.7$ and for these analyses are required to be in the
$|\eta| <$ 2.0 range. 
Jets are calibrated as a function of $\eta$ and $E_T$
and their energy is corrected to the parton level\cite{topPRD}.
Neutrinos produce missing transverse energy, $\MET$, which is measured by balancing the calorimeter 
energy in the transverse plane.
The muon sample is heavily contaminated by events produced by cosmic rays interactions with the detector.
Since these events do not originate from a common interaction vertex, 
the timing capability of the Central Outer Tracker (COT) is used to reject
events with two muon tracks, one of which travels toward the beam pipe. We also
require that the muon track passes close to the beam line, within distances less than 0.02~cm  (0.2~cm)
for tracks with (without) silicon hits.
In the analyses we are describing, 
the signal selection criteria are set according to the kinematic distribution
(e.g. $p_T$ of the muons and $E_T$ of the jets) of decay products determined from Monte Carlo studies, 
optimized to eliminate background with a minimal loss of signal events\cite{danThesis}.

In the dimuon + jets topology,
from the inclusive muon triggers dataset we select events 
with two reconstructed isolated muons with $P_T >$ 25 GeV/c. 
The first muon is required to be tight, i.e. to have a stub associated to a track,
while the second one can be without a stub (``stubless''). 
Events are further selected if there are at least two jets with
E$_T>$ 30 and 15 GeV respectively.
In the search in the muon, neutrino and two jets topology,
we select events with one reconstructed tight muon with $P_T > $ 25 GeV/c. 
We veto events with a second loose or tight muon to be orthogonal to the previous selection.
We then accept events where there is large missing transverse energy, $\MET  > 60$ 
GeV and at least two jets with $E_T>$30 GeV.

The above datasets are composed predominantly of events coming from
QCD production of $Z$/$W$  bosons in association with jets and $t\bar t$ production (where
one or both the $W$'s from top decay into  muon and neutrino).
To reduce these backgrounds we apply several cuts which depend on the final state topology. 

1) $\mu\mu$ analysis:

i) veto of events whose reconstructed dilepton mass falls in the window
$76 < m_{\mu\mu} < 110 $ GeV/c$^2$ to remove the $Z$ + jets contribution and 
$m_{\mu\mu} <$  15 GeV/c$^2$ to avoid contamination from
$J/\Psi$ and $\Upsilon$ production;
ii) $E_T(j_1) + E_T(j_2) > 85$  GeV  and  $E_T(e_1) + E_T(e_2) > 85$  GeV, 
iii) $\sqrt{ ( ( E_T(j_1) + E_T(j_2) )^2 +  (E_T(e_1) + E_T(e_2) )^2 )} >$  200 GeV.
The effect of the last two cuts is shown in Fig. 1, where SM background is
compared to a $LQ$ signal for $M_{LQ}$ = 220 GeV/c$^2$.
\begin{figure}[ht]
\begin{tabular}{c}
\includegraphics[scale=0.41]{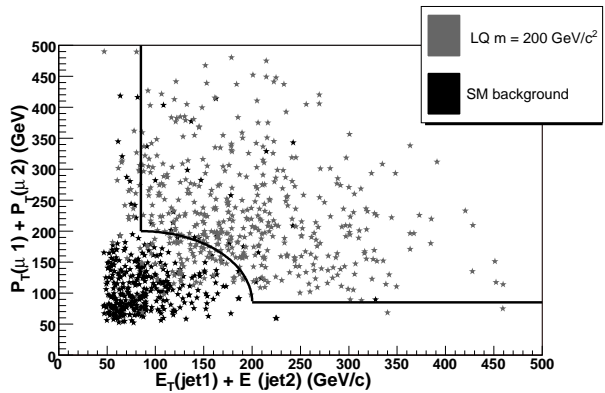} 
\end{tabular}
\caption{\label{fig:fig1}
Graphical representation of the last two topological cuts 
applied in the $\mu\mu jj$ analysis as observed on MC events.
The discrimination between SM background and $LQ$ signal is evident.}
\end{figure}

2) $\mu \nu$ analysis:

i)$\Delta\phi (\MET-jet) > 5^{\circ}$ to veto events where the 
transverse missing energy is mis-measured due to a mis-measure of the jet energy, and  
$\Delta\phi (\MET-\mu) < 175^{\circ}$ to ensure that
that the missing energy does not come from mis-measurement of the muon momentum.
ii)$E_T(j_1) + E_T(j_2) > 80$ GeV, 
iii)$M_T(\mu\nu) > 120$ GeV/c$^2$ to reduce the W + 2 jets background;
iv) a mass dependent cut consisting  
in selecting events falling in mass windows defined around several LQ masses. 
We require that the reconstructed mass combinations of the jets, muon and $\MET$ be consistent with 
those reconstructed from the LQ Monte Carlo. The ambiguity of the jet assignements allows for 
two different sets of reconstructed LQ pairs in each event, from which we build lineshapes of 
mass distributions by matching the reconstructed objects to the generator level objects.
The lineshapes are then used with the following selection:
$|M(\mu,j_1) - M_{LQ}| < 2\sigma_1$ or $|M(\mu,j_2) - M_{LQ}| < 2\sigma_2$, and 
$M_T(\MET,j_1) > T_1^{min}$ or $M_T(\MET,j_2) > T_2^{min}$,
where $M_{LQ}$ is the mean of the reconstructed LQ distribution 
and $\sigma_{1,2}$ are the width parameterizations.  
$M(\mu,j_{1,2})$ are the reconstructed muon-jet invariant masses while 
$M_T(\MET,j_{1,2})$ are the reconstructed neutrino-jet transverse 
invariant masses. 
For the transverse mass distributions, we have chosen a mass-dependent lower cut 
denoted by $T_{1,2}^{min}$: $T_1^{min} = 20 + (M_{LQ} - 120)$ GeV/c$^{2}$, and 
$T_2^{min} = 20 + (M_{LQ} - 120)/2$ GeV/c$^{2}$.
In Fig. 2 we plot the mass distributions of the selected events 
(before the mass limit cut) compared to the
signal distribution for $m_{LQ}$ = 180 GeV/c$^2$.
\begin{figure}[ht]
\begin{tabular}{c}
\includegraphics[scale=0.41]{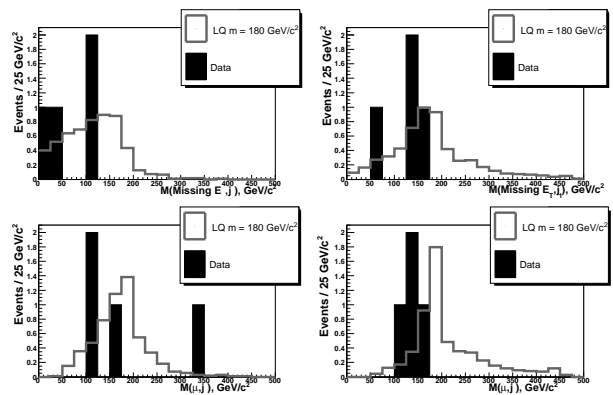} 
\end{tabular}
\caption{\label{fig:fig2}
Final mass distributions (see text) of the surviving events before the mass limit cut compared to the
signal distribution for $M_{LQ}$ = 180 GeV/c$^2$.}
\end{figure}

We study the properties of the physics backgrounds by 
generating events corresponding to $Z/W$ + 2 jets with 
ALPGEN\cite{alpgen} + HERWIG \cite{HERWIG} (to perform parton showering) 
and $t\bar t$  with PYTHIA \cite{pythia}. A  complete simulation of the CDF II detector based on 
GEANT\cite{geant} and full event reconstruction is then performed. 
To normalize the number of simulated events to data we use the theoretical cross sections for 
$t\bar t$ from \cite{mlm} and for $\gamma / Z \to \mu \mu$ + 2 jets from \cite{mcfm}.
The background arising from multijet events, where a jet is mismeasured as a muon
or where the muon comes from pion decay (QCD/fake), is evaluated using data.
In the $\mu \mu jj$ analysis we examine the data for 
same-sign events (events with two muons of the same charge) remaining after each kinematical cut. 
We estimate the background contribution to be twice the number of same
sign events, in the assumption that there is no evidence of LQ signal in these type of events (the LQ pair have
opposite charge, giving rise to two opposite charge muons).
In the $\mu \nu jj$ analysis the contribution from the QCD/fakes background is estimated by examining
the phase space of the $\MET$ vs the muon fractional isolation for data events in which the muon
isolation requirement is not enforced. Here the muon fractional isolation is defined as the ratio
between the calorimetric energy not associated with the lepton in a cone of $\Delta R$ = 0.4 
around the lepton and the energy of the lepton. 
The following assumptions are made: since jets are produced in association 
with other particles, the isolation fraction of a jet will generally be larger than the one 
corresponding to a muon; there is no correlation between the isolation of the muon and $\MET$, 
and in the region where the $\MET$ is small and the isolation of the muon is large
the LQ contribution is expected to be negligible (background-dominated region).
With these assumptions, from the ratio of the number of events in the 
background-dominated regions we can extrapolate the contribution
in the signal region.
Other backgrounds from $b\bar b$, $Z \to \tau \bar \tau$, $WW$ are negligible due to 
the muon isolation and large muon and jet transverse energy requirements. 
In the $\mu \mu$ channel the expected number of  $Z$ + 2 jets events is $1.7 \pm 0.1$.  
The expected number of $t \bar t$ events is $0.22 \pm 0.03$ events. 
We estimate $1\pm 1$ fake events 
The overall background estimate is: $3 \pm 1$ events. 
In the $\mu \nu$ channel, the number of events in each mass region, 
compared with the background expectations is reported in Table I.
\begin{widetext}
\begin{center}
\begin{table}[ht]
\caption{\label{tab:enujjfinal}
Number of events surviving all cuts in the muon, missing energy and jets topology,
compared with background expectations, as a function of the LQ mass (in GeV/c$^2$).
Errors on the expectations are both statistical and systematic.}
\begin{ruledtabular}
\begin{tabular}{cccccc}

     Mass       & 140              & 160                 & 180                 & 200                 & 220              \\
\hline 
     Wjj        & 0.9 $\pm$ 0.1    & 1.4 $\pm$ 0.1       & 1.4 $\pm$ 0.1       & 1.6 $\pm$ 0.1       & 1.6 $\pm$ 0.1    \\
     top        & 1.7 $\pm$ 0.2    & 1.8 $\pm$ 0.2       & 1.4 $\pm$ 0.2       & 1.0 $\pm$ 0.2       & 0.8 $\pm$ 0.2    \\
     Zjj        & 0.20 $\pm$ 0.01  & 0.20 $\pm$ 0.01     & 0.20 $\pm$ 0.01     & 0.20 $\pm$ 0.01     & 0.20 $\pm$ 0.01   \\
     multijets  & 0.3 $\pm$ 0.3    & 0.3 $\pm$ 0.3       & 0.3 $\pm$ 0.3       & 0.3 $\pm$ 0.3       & 0.3 $\pm$ 0.3    \\  
     Total      & 3.1 $\pm$ 0.3    & 3.7 $\pm$ 0.4       & 3.2 $\pm$ 0.3       & 3.1 $\pm$ 0.3       & 2.9 $\pm$ 0.3    \\
     Data       & 3                & 4                   & 2                   & 0                   & 0               \\
  \end{tabular}
\end{ruledtabular}
\end{table}
\end{center}
\end{widetext}
We check the prediction of our background sources with data in  
control regions where the background contribution is maximized.
For the $\mu \mu$ analysis the region is defined by requiring 
two muons with $P_T > $ 25 GeV/c,  $ 75 < m_{\mu\mu} < 105$ GeV/c$^2$ and
2 jets with $E_T >$ 30,15 GeV. We observe 110 events and expect $88\pm 10$.
For the $\mu \nu$ analysis we ask for 
one muon with $P_T > $ 25 GeV/c, $\MET >$ 35 GeV and 2 jets with $E_T >$ 30 GeV and observe 203 events
to be compared with a prediction of $221\pm 15$ from SM sources.

The efficiency to detect our signal  is obtained from 
MC simulated LQ ({\sc PYTHIA}) events to account for kinematical and geometrical acceptance. 
The total efficiencies for a LQ signal are reported in Table II.
\begin{widetext}
 \begin{center}
\begin{table}[ht]
\caption{\label{tab:eff}
Efficiencies after all cuts, errors (statistical and systematic) and 95\% C.L.
upper limits on the production cross 
section $\times$ branching fraction Br, as a function of M$_{LQ}$, for the two channels.}
\begin{ruledtabular}
\begin{tabular}{ccccc}
$M_{\rm{LQ}}$(GeV/$c^2$)  & \multicolumn{2}{c}{$\mu \mu jj$ } & \multicolumn{2}{c}{$\mu  \nu jj$  } \\
\hline
& \multicolumn{1}{l}{$\ \epsilon$} & $\sigma\times$Br(pb)
& \multicolumn{1}{l}{$\ \epsilon$} & $\sigma\times$Br(pb) \\
\hline
     100              &    0.020 $\pm$ 0.003 & 1.35          &  0.0050 $\pm$ 0.0005 &   -     \\
     120              &    0.05 $\pm$ 0.005  & 0.52          &  0.070 $\pm$ 0.005   &   0.86  \\
     160              &    0.13 $\pm$ 0.01   & 0.18          &  0.070 $\pm$ 0.005   &   0.73  \\
     200              &    0.19 $\pm$ 0.02   & 0.13          &  0.110 $\pm$ 0.005   &   0.41  \\
     220              &    0.21 $\pm$ 0.02   & 0.11          &  0.13  $\pm$ 0.01    &   0.24  \\
     240              &    0.24 $\pm$ 0.02   & 0.10          &  0.13  $\pm$ 0.01    &   0.24  \\
     260              &    0.26 $\pm$ 0.02   & 0.09          &  0.14  $\pm$ 0.01    &   0.21  \\
  \end{tabular}
\end{ruledtabular}
\end{table}
\end{center}
\end{widetext}
The following systematic uncertainties are considered when calculating 
signal acceptance and background predictions:
i) luminosity: 6\% 
ii) choice of parton distribution functions: 2.1\% 
iii) statistical error of MC $<$ 1.2\% 
iv) jet energy calibration scale  $<$ 1\%  
v) muon reconstruction : 0.8\% 
vi) $z_{vertex}$ cut : 0.5\%.
vii) initial and final  state radiation 1.8\%.
After all selection cuts, 2 events remain in the $\mu \mu$ channel, while the
number of events remaining in the $\mu \nu$ channel is reported in Table I.

In the analyses described above the number of events passing the selection cuts is consistent
with the expected number of background events. The conclusion of the two searches is that there is no LQ
signal: hence we derive an upper limit on the LQ production cross section at 95\% confidence level.
We use a Bayesian approach\cite{bayes} with a flat probability distribution for the signal cross section and  
Gaussian distributions for acceptance and background uncertainties. 
The cross section limits are tabulated in Table II and the mass limits are tabulated in Table III.
To compare our experimental results with the 
theoretical expectation, we use the next-to-leading order (NLO) cross-section for scalar LQ pair
production from \cite{kramer} with CTEQ6 parton distribution functions\cite{cteq}.
\begin{widetext}
\begin{center}
\begin{figure}[ht]
\begin{tabular}{c c}
\includegraphics[scale=0.35]{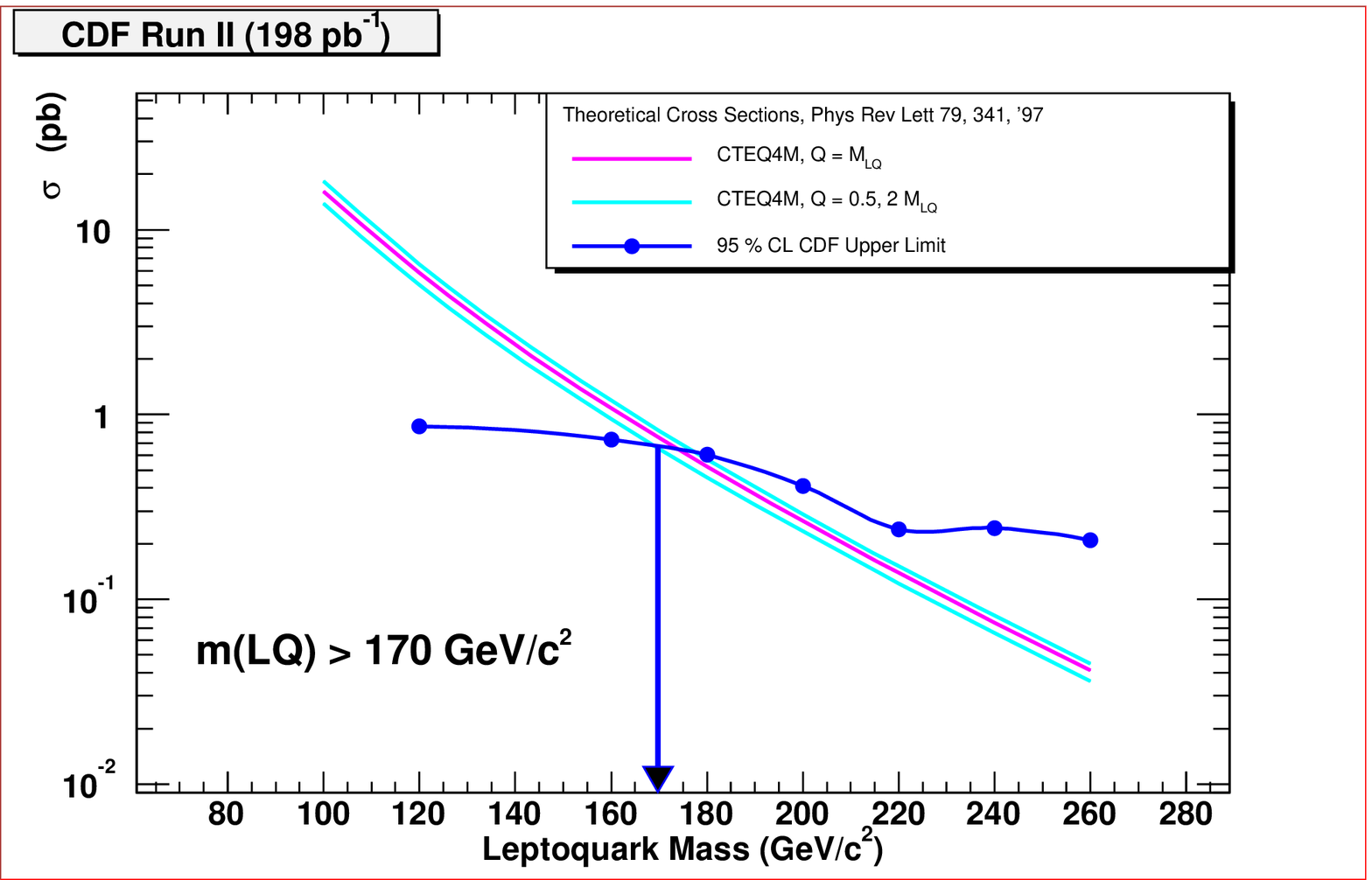} &
\includegraphics[scale=0.35]{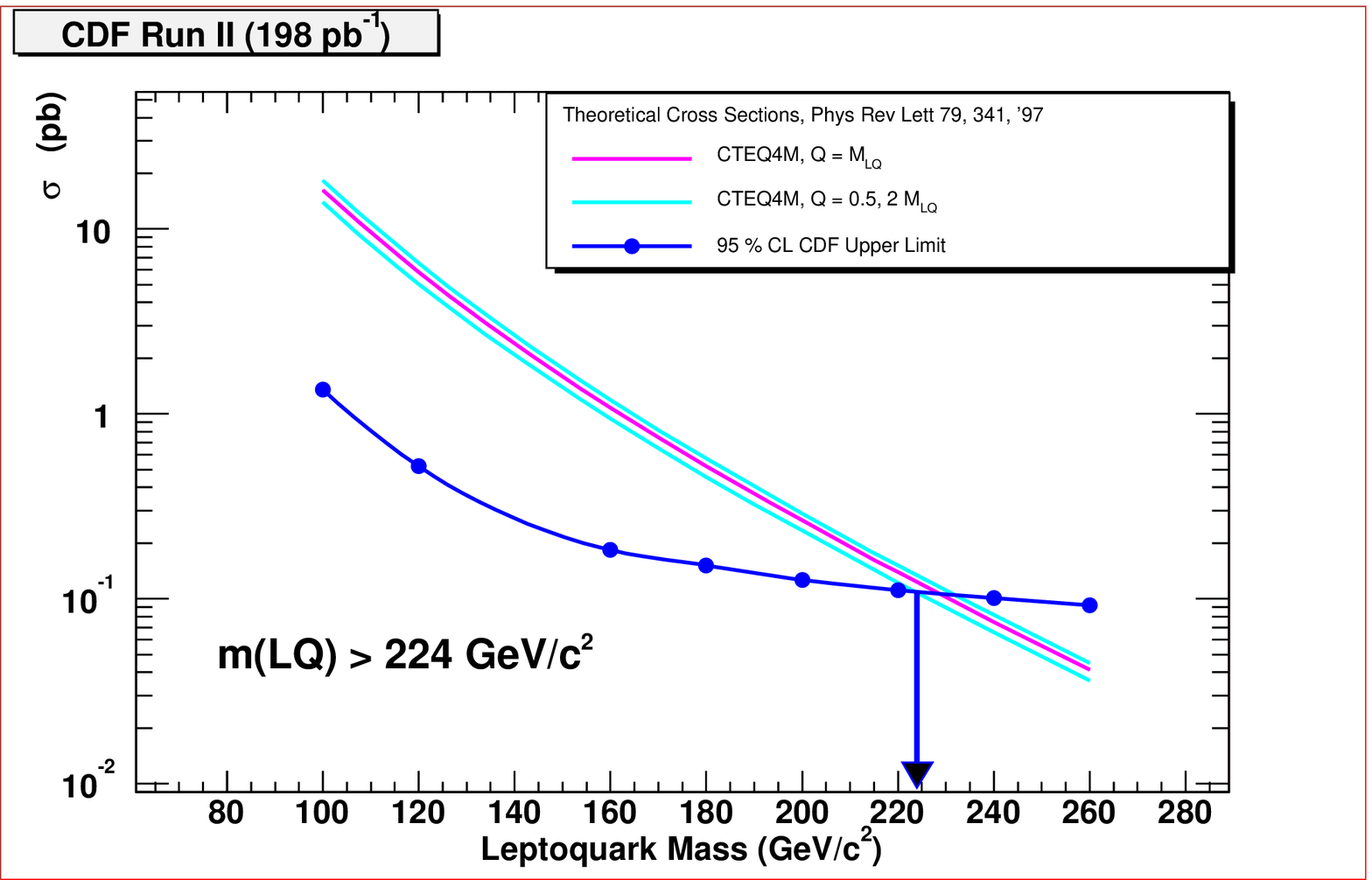}
\end{tabular}
\caption{\label{fig:fig3}
95\% C.L. limit on the experimental cross section times branching ratio as a function of
the LQ mass for the $\mu \nu jj$ and $\mu \mu jj$channel. 
The NLO theoretical cross section is plotted for different
values of the renormalization scale. Mass limits of 170 GeV/c$^2$ and 224 GeV/c$^2$ respectively are
obtained. }
\end{figure}
\end{center}
\end{widetext}
The theoretical uncertainties correspond to the variations from $M_{LQ}/2$ to $2M_{LQ}$ of the 
renormalization scale $\mu$ used in the NLO QCD calculation. To set a limit
on the LQ mass we compare our 95\% CL upper experimental limit to the theoretical cross section 
for $\mu = 2M_{LQ}$,
which is conservative as it corresponds to the lower value of the theoretical cross section.
We find lower limits on M(LQ) at 224 GeV/c$^2$ ($\beta = 1$) and 170 GeV/c$^2$ ($\beta = 0.5$).
They are reported in Figure 3.
To obtain the best limit however, we combine the results from the two decay 
channels just described  with the result of a search for LQ in the
case where the LQ pair decays to a neutrino and quark with branching ratio 
$ Br(LQ \to \nu q) = $1.0\cite{prl7200}. The individual channel analyses are in fact optimized for
fixed values of $\beta$ (1,0.5,0) while in the combined analysis, due to the contributions of the 
different decay channels, the signal acceptance can be naturally expressed as a function of $\beta$. 
As for the treatment of uncertainties, 
the searches in the $\mu \mu jj$ and $\mu \nu jj$ channels use common 
criteria and sometime apply the same kind of requirements 
so the uncertainties in the acceptances are considered correlated. When calculating the limit 
combination including the $\nu\nu jj $ channel the uncertainties are considered uncorrelated. 
For each $\beta$ value a 95\% C.L. upper 
limit on the expected number of events is returned for each mass, and by comparing this to the
theoretical expectation,  lower limits on the LQ mass are set.
The combined limit as a function of $\beta$ is shown in Figure 4, 
together with the individual channel limits. The combined mass limits are also tabulated in 
Table III.
 \begin{table}
\caption{\label{tab:finalLimit}
95\% C.L. lower limits on the second generation scalar LQ mass (in GeV/c$^2$),
as a function of $\beta$.The limit from CDF\cite{tevatron} ($\mu\mu jj$) Run I ($\sim 120 pb^{-1}$) is also given.}
\begin{ruledtabular}
\begin{tabular}{crcrcrc}
     $\beta$          & $\mu \mu$ jj  & $\mu \nu jj$  &  $\nu \nu jj$   &   Combined  & CDF Run I \\
     0.01             & -      &  -          & 114             &   125       & -\\
     0.05             & -      &  -          & 110             &   133       & - \\
     0.1              & -      &  137        & -               &   143       & - \\
     0.2              & -      &  155        & -               &   157       & - \\
     0.3              & 100    &  162        & -               &   176       & - \\
     0.4              & 152    &  168        & -               &   200       & - \\
     0.5              & 171    &  170        & -               &   208       & - \\
     0.6              & 184    &  168        & -               &   213       & - \\
     0.7              & 196    &  162        & -               &   217       & - \\
     0.8              & 206    &  155        & -               &   221       & - \\
     0.9              & 215    &  137        & -               &   224       & - \\
     1.0              & 224    &  -          & -               &   226       & 202 \\
  \end{tabular}
\end{ruledtabular}
\end{table}
\begin{figure}[h]
\includegraphics[scale=0.41]{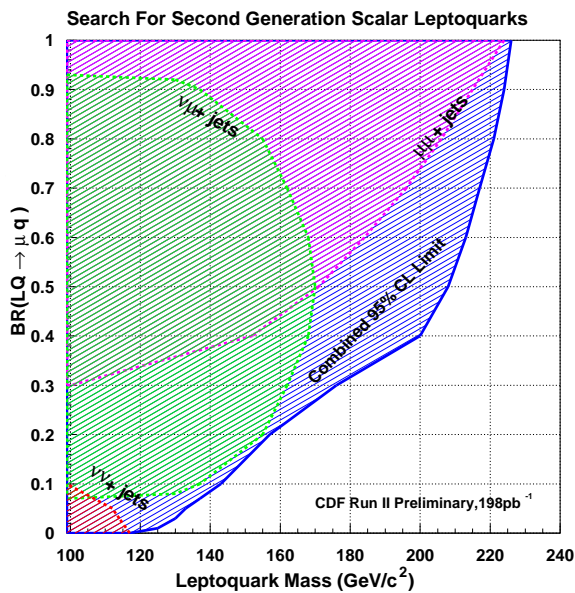} 
\caption{\label{fig:limit_combined}
Leptoquark mass exclusion regions at 95\% C.L. as function of Br(LQ$\to  \mu$ q).}
\end{figure}

The final result presented here is better than the results obtained with Tevatron Run I 
data\cite{SecGenTeV1}. 
This is mostly due to the small increase in the cross section as a function of the center of mass energy
(from 1.8 to 1.96 TeV), and an increase in the muon acceptance. 
A comment is in order when comparing this result with the ones recently published\cite{newTeV} for
first generation LQ and third generation LQ\cite{tau}. 
While the signatures of LQ production
is very similar (high $P_T$ leptons, large transverse missing energy and energetic jets) one has to consider
the constraint on the LQ particle to only couple to same generation fermions. This implies different types of
selection and exclude the possibility of combining inter-generation results.
Also, in the case of electrons and muons, as can be seen from the current results, the similarity in their
acceptance results in similar cross section and mass limits, while the result is quite different in the
case of third generation LQ, due to the much smaller acceptance for $\tau$ leptons.
The results presented in this paper, as well as the ones recently
published in \cite{newTeV}, are obtained with a statistical sample corresponding
to about twice the luminosity collected in Tevatron Run I. A future
increase in the Run II luminosity by an order of magnitude is estimated to extend the LQ mass
range to $\sim$ 300 GeV/c$^2$ for the $\beta$ = 1 case \cite{simonaTeV4LHC}. Substantially higher LQ
masses will be explored at the future Large Hadron Collider (LHC)\cite{zerwas}.

In conclusion, we have performed a search for pair production of second generation
scalar LQ in  the dimuons + jets and muon, missing energy + jets topologies, using 
198 pb$^{-1}$ of proton-antiproton
collision data recorded by the CDF experiment during Run II of the Tevatron. 
We combined these findings with the 
ones from a search in the $\MET$ + jets topology\cite{prl7200}. No evidence for LQ is observed.
Assuming that the LQ decays to muon and quark with variable branching ratio $\beta$
we exclude LQ with masses below 226 GeV/c$^2$ for $\beta$ = 1, 208 GeV/c$^2$ for $\beta$ = 0.5 
and 143 GeV/c$^2$ for $\beta$ = 0.1 at 95\% C.L.

We thank the Fermilab staff and the technical staffs of the participating
institutions for their vital contributions. This work was supported by the
U.S. Department of Energy and National Science Foundation; the Italian
Istituto Nazionale di Fisica Nucleare; the Ministry of Education, Culture,
Sports, Science and Technology of Japan; the Natural Sciences and
Engineering Research Council of Canada; the National Science Council of the
Republic of China; the Swiss National Science Foundation; the A.P. Sloan
Foundation; the Bundesministerium fuer Bildung und Forschung, Germany; the
Korean Science and Engineering Foundation and the Korean Research
Foundation; the Particle Physics and Astronomy Research Council and the
Royal Society, UK; the Russian Foundation for Basic Research; the  Comisi\'on
Interministerial de Ciencia y Tecnolog\'{\i}a, Spain; and in part by the
European Community's Human Potential Programme under contract HPRN-CT-20002,
Probe for New Physics.

\end{document}